\documentclass{article}

\usepackage{arxiv}

\usepackage[utf8]{inputenc} 
\usepackage[T1]{fontenc}    
\usepackage{hyperref}       
\usepackage{url}            
\usepackage{booktabs}       
\usepackage{amsfonts}       
\usepackage{nicefrac}       
\usepackage{microtype}      
\usepackage{lipsum}
\usepackage{graphicx}
\usepackage{float}

\usepackage{booktabs}
\usepackage{multirow}

\graphicspath{ {./images/} }

\title{Towards Automatic building of Human-Machine Conversational System to support Maintenance Processes}

\author{
 Elena Coli \\
  Department of Information Engineering\\
  University of Pisa\\
  \texttt{elena.coli@phd.unipi.it}
   \And
 Nicola Melluso \\
  Department of Energy Systems, Territory and Construction Engineering\\
  University of Pisa\\
  \texttt{nicola.melluso@phd.unipi.it}
  \And
 Gualtiero Fantoni \\
  Department of Civil and Industrial Engineering\\
  University of Pisa
  \And
 Daniele Mazzei \\
  Department of Computer Science\\
  University of Pisa
}

\begin{document}
\maketitle

\begin{figure}[H]
  \centering
  \includegraphics[keepaspectratio=true,scale=0.2]{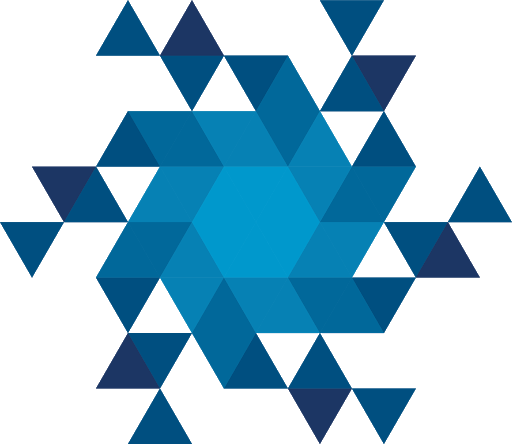}
\end{figure}

\begin{abstract}
Companies are dealing with many cognitive changes with the introduction of the Industry 4.0 paradigm. In this constantly changing environment, knowledge management is a key factor. Dialog systems, being able to hold a conversation with humans, could support the knowledge management in business environment. Although, these systems are currently hand-coded and need the intervention of a human being in writing all the possible questions and answers, and then planning the interactions. This process, besides being time-consuming, is not scalable. Conversely, a dialog system, also referred to as chatbot, can be built from scratch by simply extracting rules from technical documentation. So, the goal of this research is designing a methodology for automatic building of human-machine conversational system, able to interact in an industrial environment.
An initial taxonomy, containing entities expected to be found in maintenance manuals, is used to identify the relevant sentences of a manual provided by the company BOBST SA and applying text mining techniques, it is automatically expanded. The final result is a taxonomy network representing the entities and their relation, that will be used in future works for managing the interactions of a maintenance chatbot.
\end{abstract}


\section{Introduction}
In literature, the concept of “Industry 4.0” is often linked to its nine enabling technologies, defined for the first time by the Boston Consulting Group [1]. Although, many works underline that Industry 4.0 is not only a mere application of these technologies, but also involves many organizational and management challenges to better face the competition [2]: technologies are meaningful if the company are able to extract the real value with workers able to use and maintain them. Thus, knowledge management becomes crucial for companies facing the Fourth Industrial Revolution.

Knowledge management could be supported by the implementation of dialog systems, able to spread knowledge within an organization. But actually, dialog systems are hand-coded [3], first identifying the needed functions, and then planning the interactions with the users. Therefore, the building of a chatbot is a time-consuming and not scalable process.

The research goal is to design a methodology for automatic building of human-machine conversational system, able to interact in an industrial environment. The aim of this paper is to provide a first step towards this objective, by automatically capturing the knowledge underpinning a maintenance process, in order to build knowledge base that will be the input for the conversational system.

Since the methodology is based on text mining techniques, an initial maintenance taxonomy, containing entities (such as components, verbs etc.) likely to be found in a maintenance manual, is used to identify relevant sentences in a technical document provided by the company BOBST SA. Then, the taxonomy is automatically expanded using these sentences and the main result is a taxonomy network, representing the entities and their relations.

This paper is structured as follow: in section 2, a literature review is carried out for helping the reader to deeply understand both the importance of the topic of knowledge in the digital age and the link between knowledge management and conversational systems. Moreover, an overview on the tools underpinning the mapping of the knowledge repository of a chatbot is provided. Then, the designed methodology and its application are explored in section 3 and the main results of the research are outlined in section 4. In the end, in section 5, the comments of the authors about the results are remarked and the future developments of the work are highlighted.

\section{Literature Review}
\subsection{Knowledge Management in Industry 4.0}
\label{sec:knowmang}
Knowledge in an organization is the collection of expertise, experience, and information that individuals and workgroups use during the execution of their tasks [4]. Even if in literature there are many taxonomies and models for describing knowledge from different perspectives, the most relevant distinction is between tacit and explicit knowledge. Tacit knowledge is embedded in people mind and it is difficult, if not impossible, to be exploited [5]. Explicit knowledge exists as text documents, structured databases, images and many other forms. For this reason, it is easier to be formalized and, consequently, shared within an organization.

Despite the fact that philosophers, scientists and writers have been wondering for centuries about how to create, acquire, communicate knowledge and, in particular, how to re-use it, only in the last 25-30 years knowledge management has been universally recognized as a self-sustaining research topic [6].
Currently, the knowledge management could be described by using different models. In particular, the model we have taken into account for this research consists of six core phases [6]: generate, refine, store, transfer, share and use knowledge. Knowledge management’s main objective is to improve each one of the six steps.

The knowledge management becomes even more relevant in the context of Industry 4.0. I4.0 is defined as a trend of automation that differentiates itself from previous industrial paradigms because of its global scope, its exponential growth and its still uncertain (but for sure powerful) impact [7]. People talking about this new paradigm usually focus on its enabling technologies (such as 3D printing and clouds), leaving aside the role that data and information play in the digital age. Actually, data are hidden behind each one of the technologies 4.0: just think about the already mentioned clouds, which allow the storage and transmission of data, or the simulation, which is made possible only by the availability and modeling of data. This thesis is confirmed also by the Acatech study on the new “Industrie 4.0” [8], that recognizes knowledge management as one of the missing building blocks of the Fourth Industrial Revolution.

Therefore, the most revolutionary aspect of I4.0 is not acquiring new machines but being able to manage the knowledge needed to take full advantage from them. I4.0 phenomenon leads the workers to solve non-standardized problems by using their knowledge: in this, workers could be supported by many technologies 4.0 for generating, refining, storing, transferring, sharing and use (and re-use) knowledge [9]. The concept of “knowledge worker”, the man or woman who applies to productive work ideas, concepts, and information rather than manual skill or brawn [10], and, consequently, the concept of “knowledge management” are definitely enhanced by the Fourth Industrial Revolution. 

\subsection{Conversational systems to manage knowledge}
In literature, it is recognized that the knowledge sharing could be supported by the implementation of a query system and routing the queries to a knowledge expert [11]. Although, the success of this kind of approach is strongly dependent from human intervention and a bottleneck arises when the limit of queries that can be processed by a human being is reached. This problem could be solved by implementing a conversational system that is potentially able to answer to an infinite number of queries coming from any locations [12]. In fact, many researches on knowledge management, as for example the one conducted by Schacht and Mädche [13], provide not only mechanisms for direct communication with experts (e.g. social networking sites, forums, chats), but also design principles mainly focused on the externalization of knowledge, its storage, retrieval and reuse.

Conversational systems could be classified in two categories: chatbots and dialog systems. The main differences between the two ones are that dialog systems (1) are usually built to be used in a more specific domain than chatbots and (2) have a more complex architecture than chatbot [14]. In fact, while chatbots consist only in a set of predefined responses and a pattern matching module, dialog systems consist of four modules, each one with a different function (pre-processing, natural language understanding, dialog managing and response generating) [15]. However, even in literature the two categories are not clearly distinct, so we can consider the two expressions mostly interchangeable and having the same meaning: machines able to hold a conversation with another agent or with a human [16].

Such systems are increasingly gaining ground in the consumer market (for example Amazon Echo, which incorporates the intelligence of the virtual assistant Alexa). Moreover, they are upsetting the communication between companies and customers, too. Many cases of chatbot applications for customer care and e-commerce management could be identified. Although, chatbot could be used in many other scenarios: one of the most interesting applications is in supporting workers in their everyday tasks. At the same time, implementing chatbots in business scenarios represents a big challenge: in fact, they still encounter some resistance in industrial contexts, revealing that the conversational systems designed for the workplace are not achieving the same success of the other ones.

Regardless of the application field (business or consumer), the brain of the chatbot is the knowledge base in which the possible responses are contained: on the basis of the user input, the dialog managing module is able to generate the most likely matching output. The main issue of the knowledge bases currently used in chatbots is that they are hand-coded. Therefore, the building of chatbot knowledge base is time-consuming and difficult to adapt to different cases and domains [3]. There are many works aimed to automatically build a chatbot knowledge base, among these we can mention the one conducted by Shawar and Atwell [17]. However, as far as we know, these works use as key tool the human annotation: therefore, they cannot solve the problem of time-consuming human intervention.

The intuition behind this research is that, conversely, the knowledge base of a dialog system for industrial applications can be built from scratch by simply extracting rules from technical documentation. Such rules will be then fed to an engine (expert system) suitable to manage the chatbot. This makes the chatbot building independent from the human intervention and, consequently, faster and cost-efficient. The interest in this direction is confirmed by many previous works: Ramesh et al. [18] highlight that chatbots have now come a long way from simple retrieval-based pattern matching approach to deep learning based neural networks one.

The extraction rules could be applied to every kind of technical documentation: technical specifications, standards, quotations, user manuals etc. The focus of this research is on the building of a chatbot to support maintenance processes, so the work involves the extraction of information from a specific type of technical document: the maintenance manual. The reason behind this choice is giving a strong priority to maintenance topic. In fact, maintenance assumes a key role in I4.0 context: with the increasing complexity, scope, and organisational role of new technologies 4.0, their maintenance is becoming a critical factor for determining the competitive success of an organization [19].

\subsection{Knowledge Mapping for Conversational Systems}
The building process of such conversational systems able to interact in an industrial maintenance process environment, relies upon the retrieval of the effective value-added information stored in the process documents.

These chatbots must be able to connect human speech with the intent they were equipped and helping an operator to find answers about a particular process or machine. These systems owe their skill to Natural Language Processing (NLP) that examines human speech and makes use of knowledge about the sentences structure, extracting entities about the maintenance process using machine-learned pattern recognition. So that entities, nothing more than data buckets used in conversation, are the building blocks of the knowledge base that run over the conversational system and extracting them to fed this body of knowledge becomes of crucial importance.

In general, Named Entity Recognition (NER) is the task of identifying entity names like people, organizations, places, temporal expressions or numerical expressions [20]. Methods and algorithms to deal with the entity extraction task are different, and since the most effective are the ones based on supervised methods, our dealt with a very closed domain moves our approach through an automatic extraction that focuses on evaluation sentence-oriented statistics of individual word [21].
In the very first step of this line of research, we focus on the development of such entities and their relations that occur in the maintenance process as representation of knowledge that will be the input for designing and building our conversational system.

\section{Materials and Methods}
For reaching the goal, a dedicated workflow has been designed (Figure 1).

\begin{figure}[H]
  \centering
  \includegraphics[keepaspectratio=true,scale=0.45]{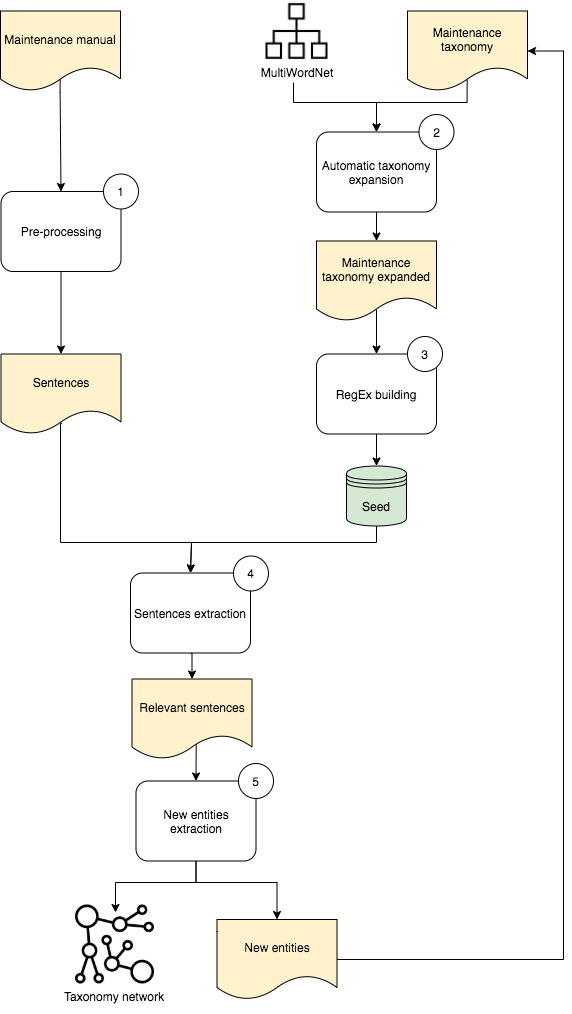}
  \caption{Research Workflow}
  \label{fig:fig1}
\end{figure}

Firstly, the maintenance manual is pre-processed, obtaining a list of sentences (1). Concurrently, a taxonomy of entities expected to be found in a maintenance manual (such as verbs, components, DPI, etc.) is automatically expanded (2) through synonyms extracted using MultiWordNet [22]. The output is an expanded maintenance taxonomy. Then, regular expressions associated to the entities are set up, building the seed, a collection of elements useful for running the extraction process (3). After that, the seed is used for the extraction of relevant sentences, containing at least one of its elements (4). New entities are automatically extracted from the relevant sentences (5), by using an automatic keyword extraction algorithm [21]. Then, the extracted entities are manually checked in order to validate them and the taxonomy is expanded with the new approved entities, for the process improvement in further applications. Finally, the entities and their relation are represented in a taxonomy network.

In this paper, we ran an experiment using as a source a maintenance manual provided by BOBST SA, a global company that produces presses for rotogravure printing and coating and laminating machines for the flexible materials industry. The manual describes the maintenance operations of a flexographic printing machine for labels and flexible packaging and it is written in Italian.
The results of method application will be deeply described and shown in the following sections.

\subsection{Preprocessing}
The work started with the pre-processing of the maintenance manual. The manual was a pdf document: firstly, it was converted in a plain text file in order to be processed. Then, the text was splitted in single sentences.

\begin{figure}[H]
  \centering
  \includegraphics[keepaspectratio=true,scale=0.6]{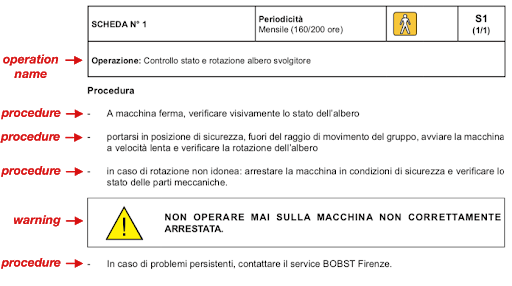}
  \caption{Extract from a page of BOBST manual}
  \label{fig:fig2}
\end{figure}

This manual consists of many pages, each one containing a maintenance card corresponding to a single maintenance operation (Figure \ref{fig:fig2}).

This implies that all the pages have the same structure. This reasoning could be applied to every maintenance manual: each manual is almost structured and, once identified the recurring structures, we can easily understand its recurring contents. For example, the procedures are always expressed by using bullet points. Based on these criteria, the typology of sentence was automatically identified, among the following:
\begin{itemize}
    \item operation names;
    \item materials;
    \item procedures;
    \item preliminary notes and warnings;
    \item other (not relevant elements).
\end{itemize}

This assignment is relevant not only for the next phase of entities extraction, but also for the setting of conversational system interaction (in fact, the chatbot will be asked to “understand” if it is communicating a procedure rather than a warning).
Table \ref{tab:tab1} shows some example of splitted sentences and associated sentence types.

\begin{table}[H]
\caption{Extract from splitted sentences and associated types}
\label{tab:tab1}
\centering
\begin{tabular}{lll}
\hline
\textbf{sentence ID} & \textbf{sentence type}  & \textbf{sentence}                                                                                                                                                                                                   \\ \hline
\textit{00001OTH4}   & \textit{other}          & Istruzioni originali                                                                                                                                                                                                \\
\textit{00001OTH5}   & \textit{other}          & © BOBST 2018                                                                                                                                                                                                        \\
\textit{00025OTH1}   & \textit{operation name} & Controllo impianto pneumatico                                                                                                                                                                                       \\
\textit{00025PRC1}   & \textit{procedure}      & \begin{tabular}[c]{@{}l@{}}Verificare l’integrità del circuito e, in caso di \\ necessità, sostituire le parti non integre\end{tabular}                                                                             \\
\textit{00025PRC2}   & \textit{procedure}      & \begin{tabular}[c]{@{}l@{}}a macchina arrestata, provare il \\ funzionamento dei pulsanti pneumatici\end{tabular}                                                                                                   \\
\textit{00025PRC3}   & \textit{procedure}      & \begin{tabular}[c]{@{}l@{}}verifica del circuito: a macchina arrestata, \\ rimuovere i ripari posteriori e procedere al \\ controllo dei vari attacchi pneumatici\end{tabular}                                      \\
\textit{00025PRC4}   & \textit{procedure}      & \begin{tabular}[c]{@{}l@{}}esaminare lo stato del sistema di alimentazione \\ pneumatica: stato dei tubi, stato dei sistemi di \\ innesto rapido; in caso di necessità intervenire \\ con sostituzioni\end{tabular} \\
\textit{00033OPR1}   & \textit{note/warning}   & \begin{tabular}[c]{@{}l@{}}NON UTILIZZARE CREME O POLVERI \\ ABRASIVE, NON UTILIZZARE DETERGENTI \\ AGGRESSIVI O ACIDI.\end{tabular}                                                                                \\
\textit{00033PRC1}   & \textit{procedure}      & \begin{tabular}[c]{@{}l@{}}Procedere a periodica e metodica \\ pulizia per mantenere in buono stato \\ la macchina\end{tabular}                                                                                     \\
\textit{00035OTH1}   & \textit{operation name} & Controllo generale sistema di trasmissione                                                                                                                                                                          \\
\textit{00035PRC1}   & \textit{procedure}      & \begin{tabular}[c]{@{}l@{}}Verificare lo stato del sistema e l’efficienza \\ del sistema di trasmissione per garantire \\ il corretto funzionamento della macchina\end{tabular}                                     \\
\textit{00037PRC1}   & \textit{procedure}      & \begin{tabular}[c]{@{}l@{}}Smontare la vaschetta e provvedere alla \\ periodica pulizia con detergente non aggressivo\end{tabular}                                                                                  \\
\textit{00037PRC2}   & \textit{procedure}      & \begin{tabular}[c]{@{}l@{}}periodicamente, e in via straordinaria, ogni \\ qualvolta si rilevino anomalie nella stampa, \\ occorre provvedere alla pulizia dei supporti\end{tabular}                                \\
\textit{00037PRC3}   & \textit{procedure}      & \begin{tabular}[c]{@{}l@{}}smontare il supporto agendo sul bullone \\ di fissaggio\end{tabular}                                                                                                                     \\
\textit{00037PRC4}   & \textit{procedure}      & \begin{tabular}[c]{@{}l@{}}pulire il supporto avendo cura di non \\ modificare la regolazione del dado e del \\ controdado del perno con molla\end{tabular}                                                         \\
\textit{00063MTR1}   & \textit{material}       & Aspiratore                                                                                                                                                                                                          \\
…                    & …                       & …                                                                                                                                                                                                                   \\ \hline
\end{tabular}
\end{table}

We can state that in the technical domain of maintenance, texts (such as manuals) follow not only the same page architecture, but also the same language formalism. This implies that different sentence types follow different recurring grammatical structures (for example, the procedures contain only verbs in the infinitive form). So, we can be confident that, once identified these recurring grammatical structures, the sentence extraction will not be subject to ambiguity issues and will be replicable without human intervention.

\subsection{Automatic taxonomy expansion}

Concurrently, a maintenance taxonomy was manually generated. Because of the wide variety of entities expected to be found in a maintenance manual, it is crucial to organize them in classes. In the specific maintenance domain, the following classes were identified:
\begin{itemize}
    \item advantage: verbs, nouns and adjectives describing an advantage for the user;
    \item alert: words giving an alert to the user;
    \item chemistry: chemical elements;
    \item chemistry: chemical elements;
    \item component: machine components;
    \item DPI: personal protective equipements;
    \item drawback: verbs, nouns and adjectives describing a drawback for the user;
    \item lifecycle: phases of machine life cycle;
    \item math: mathematical symbols and expressions;
    \item qualification: qualifications of the operators;
    \item tool: tools to be used during operations;
    \item verb: verbs.
\end{itemize}

The identification and population of the classes are due to this specific manual and, of course, the analysis of additional maintenance manuals could lead to the identification of further classes and entities. These two phases were performed by using both the expertise of domain technicians and the results of previous research works (such as advantages and drawbacks [23] and functional verbs [24]). 1271 entities were identified, stored and classified in the maintenance taxonomy.

The taxonomy was then automatically expanded. In fact, due to the large number of ways to express a concept in Italian, it has proved necessary to identify entities synonyms. Synonyms were identified by using MultiWordNet, reaching an expanded taxonomy of 3943 entities.
Table \ref{tab:tab2} shows an extract from the maintenance taxonomy expanded, containing entities and associated synonyms and classes.

\begin{table}[H]
\caption{Extract from maintenance taxonomy expanded}
\label{tab:tab2}
\centering
\begin{tabular}{lll}
\hline
\textbf{class} & \textbf{entity} & \textbf{synonyms}                                                                                                                                                                                                                       \\ \hline
advantages     & garantire       & \begin{tabular}[c]{@{}l@{}}promettere, appurare, assodare, \\ guardare, assicurare, controllare, \\ vedere, garentire\end{tabular}                                                                                                      \\
alert          & attenzione      & \begin{tabular}[c]{@{}l@{}}gentilezza, cautela, circospezione, \\ avvedutezza, cortesia, amorevolezza, \\ cura, dolcezza, atto\_di\_cortesia, \\ premura, riguardo, prudenza, avvertenza, \\ precauzione, favore, carità\end{tabular}   \\
chemistry      & Tecnezio        & tecneto                                                                                                                                                                                                                                 \\
component      & martinetto      & cricco, binda, martinello                                                                                                                                                                                                               \\
DPI            & elmo            & elmetto, cimiere, cimiero, casco                                                                                                                                                                                                        \\
drawbacks      & insufficienza   & \begin{tabular}[c]{@{}l@{}}carestia, penuria, minimezza, deficienza, \\ carenza, pochezza, deficit, spareggio, \\ limitatezza, ristrettezza, magrezza, \\ esiguità, sparutezza, povertà, pocanza, \\ disavanzo, meschinità\end{tabular} \\
drawbacks      & insufficienza   & \begin{tabular}[c]{@{}l@{}}minimezza, deficienza, carenza, pochezza, \\ deficit, spareggio, limitatezza, magrezza, \\ esiguità, sparutezza, povertà, pocanza, \\ disavanzo, meschinità\end{tabular}                                     \\
math           & \textgreater{}  &                                                                                                                                                                                                                                         \\
qualifica      & modellista      & disegnatore\_di\_moda, couturier, stilista                                                                                                                                                                                              \\
utensile       & cesoie          & trancia, forbici\_da\_tosatore, tranciatrice                                                                                                                                                                                            \\
verb           & Abbassare       & \begin{tabular}[c]{@{}l@{}}ribassare, incupire, incupirsi, scurire, \\ chinare, diminuire, tagliare, scurirsi, ridurre, \\ abbassare, inchinarsi, assottigliare, \\ amputare, inchinare\end{tabular}                                    \\
vita           & declino         & \begin{tabular}[c]{@{}l@{}}decadenza, deperimento, deterioramento, \\ peggioramento, scadimento, aggravamento, \\ degrado, deteriorazione, decadimento\end{tabular}                                                                     \\
\\
…                    & …                       & … \\ \hline
\end{tabular}
\end{table}

\subsection{RegEx Building}

In order to perform the sentences extraction, it is necessary to translate the human language in a language that the machine is able to use in order to perform the extraction task. This is possible by using regular expressions (also called regEx), sequences of characters that define a search pattern [25]. One regular expression for each class was built. The output of the phase is the seed, a collection of elements useful for running the extraction process, whose structure is shown in the extract in Table \ref{tab:tab3}.

\begin{table}[H]
\centering
\caption{Extract from the seed}
\label{tab:tab3}
\begin{tabular}{ll}
\hline
\textbf{class} & \textbf{regex}                                                                                                                                                                                                                                                                                                                                                                                                                                                                                                                                                                                                                                                                                                                                                                                                                                                                                                                                                                                                                                   \\ \hline
DPI            & \begin{tabular}[c]{@{}l@{}}(girella)|(carrucola)|(puleggia)|(cimiere)|(essiccatoio)|\\ (asciugatoio)|(cimiero)|(elmo)|(essiccatore)|(casco)|\\ (elmetto)|(acconciatura)|(cappello)|(copricapo)|(elmetto)|\\ (casco)|(elmo)|(cimiero)|(cimiere)|(elmo)|(casco)|\\ (elmetto)|(cimiere)|(cimiero)|(imbragatura)|(imbracatura)|\\ (imbragatura)|(imbracatura)|(inganno)|(manto)|(maschera)|\\ (pretesto)|(puntale)|(punta)|(schermo)|(copertura)|(video)|\\ (visiera)|(buffa)\\  \\\end{tabular}                                                                                                                                                                                                                                                                                                                                                                                                                                                                                                                                                           \\
alert          & \begin{tabular}[c]{@{}l@{}}(avvertenza)|(attenzione)|(carità)|(favore)|(precauzione)|\\ (prudenza)|(premura)|(riguardo)|(avvedutezza)|\\ (circospezione)|(cortesia)|(gentilezza)|(cautela)|\\ (amorevolezza)|(cura)|(dolcezza)|(atto\_di\_cortesia)\\  \\\end{tabular}                                                                                                                                                                                                                                                                                                                                                                                                                                                                                                                                                                                                                                                                                                                                                                                 \\
qualification  & \begin{tabular}[c]{@{}l@{}}(couturier)|(stilista)|(modellista)|(disegnatore\_di\_moda)|\\ (sensale)|(corrispondente)|(cozzone)|(prosseneta)|\\ (agente)|(rappresentante)|(operatore)|(acquisitore)|\\ (cineoperatore)|(cameraman)|(corrispondente\_commerciale)|\\ (manovratore)|(intermediario)|(riparatore)|(racconciatore)\\  \\\end{tabular}                                                                                                                                                                                                                                                                                                                                                                                                                                                                                                                                                                                                                                                                                                       \\
lifecycle      & \begin{tabular}[c]{@{}l@{}}(crescita)|(espansione)|(evoluzione)|(aumento)|(ontogenesi)|\\ (aggravio)|(embriogenesi)|(sviluppo)|(decadimento)|\\ (scadimento)|(deteriorazione)|(degrado)|(aggravamento)|\\ (deterioramento)|(deperimento)|(declino)|(decadenza)|\\ (peggioramento)|(slargamento)|(allargamento)|(allargatura)|\\ (espandimento)|(espansione)|(ampliamento)|(sviluppo)|\\ (estensione)|(crescita)|(proemio)|(inserimento)|(presentazione)|\\ (prolusione)|(prefazione)|(isagoge)|(cappello)|(avvertimento)|\\ (inserzione)|(introduzione)|(maturità)|(età\_adulta)|(maturazione)|\\ (impostare)|(insegnare)|(progresso)|(avanzamento)|(istruire)|\\ (nascere)|(passo\_avanti)|(formulare)|(ontogenesi)|(sviluppare)|\\ (formare)|(espansione)|(evoluzione)|(espansione)|(esercitare)|\\ (crescita)|(sorgere)|(embriogenesi)|(progresso)|(passo\_in\_avanti)|\\ (ontogenesi)|(crescita)|(avanzamento)|(evoluzione)|(passo\_avanti)|\\ (educare)|(embriogenesi)|(derivare)|(allenare)|\\ (passo\_in\_avanti)|(sviluppo)\end{tabular}

\\
…                    & …                       \\ \hline
\end{tabular}
\end{table}

\subsection{Sentences Extraction}

The seed was used for identifying, among the sentences, only those containing at least one of its elements. The extracted sentences are those more relevant to the maintenance topic.
The phase allowed the extraction of 151 relevant sentences. Table \ref{tab:tab4} 6 shows 3 examples of extracted sentences and associated ID, along with the matching entity and associated class.

\begin{table}[H]
\centering
\caption{Extract from relevant sentences}
\label{tab:tab4}
\begin{tabular}{@{}llll@{}}
\toprule
\textbf{sentence ID}       & \textbf{sentence}                                                                                                                                                                                                                                                  & \textbf{entity matched} & \textbf{entity class} \\ \midrule
\multirow{3}{*}{00015PRC2} & \multirow{3}{*}{\begin{tabular}[c]{@{}l@{}}rimuovere il riparo posteriore in \\ corrispondenza del gruppo \\ svolgitore e verificare manualmente \\ lo stato di tensione della \\ cinghia di trasmissione \\  \\ \end{tabular}}                                            & component               & cinghia               \\
                           &                                                                                                                                                                                                                                                                    & verb                    & rimuovere             \\
                           &                                                                                                                                                                                                                                                                    & verb                    & verificare            \\
\multirow{4}{*}{00008OPR1} & \multirow{4}{*}{\begin{tabular}[c]{@{}l@{}}\\ \\ nel caso in cui si renda necessario \\ smontare griglie o carter di protezione, \\ al termine dei lavori di manutenzione \\ è necessario procedere al loro \\ rimontaggio, prima di riavviare la macchina \\ \\ \\\end{tabular}} & drawbacks               & cedere                \\
                           &                                                                                                                                                                                                                                                                    & verb                    & smontare              \\
                           &                                                                                                                                                                                                                                                                    & verb                    & procedere             \\
                           &                                                                                                                                                                                                                                                                    & verb                    & riavviare             \\
\multirow{5}{*}{00069PRC3} & \multirow{5}{*}{\begin{tabular}[c]{@{}l@{}}\\ \\ \\ smontare il cuscinetto di supporto (2),\\  dopo aver rimosso l’anello \\ \\\end{tabular}}                                                                                                    & advantages              & supporto              \\
                           &                                                                                                                                                                                                                                                                    & component               & cuscinetto            \\
                           &                                                                                                                                                                                                                                                                    & drawbacks               & arresto               \\
                           &                                                                                                                                                                                                                                                                    & utensile                & pinza                 \\
                           &                                                                                                                                                                                                                                                                    & verb                    & smontare              \\ \bottomrule
\end{tabular}
\end{table}

\subsection{New Entities Extraction}

The results of the previous phase have highlighted the need to further expand the taxonomy. In fact, as could be seen in Table \ref{tab:tab4}, the relevant sentences contain, as well as the entities of the seed, many others not considered but relevant for the building of an expert system able to manage a conversation.
The RAKE algorithm [21] permits the automatic extraction of those new entities. As parameter, it has been decided to set to 2 the maximum number of words composing the entities.
After the extraction, the new expressions were manually checked in order to validate each one of them before updating the taxonomy. In case of positive check, the entities are used to:
\begin{enumerate}
    \item expand the existing classes;
    \item create new classes, in case the entity does not match an already existing one.
\end{enumerate}

Since we can assume that, for different manuals, we could have a common list of rejected entities from future new entities extractions, it has been decided to include the rejected entities in a blacklist, namely a list of entities not to be considered for the taxonomy expansion. The blacklist could reduce the complexity of future checking phase. Although, it has to be considered that this approach could introduce some biases, because the rejected entities are strongly linked not only to the domain (in this case maintenance), but also to other variables (for example the typology of machine).

In total, 505 new entities were identified.
During the checking process, besides the blacklist, 10 new classes were created:

\begin{itemize}
    \item \textit{operation}: nouns (not verbs) describing operations;
    \item \textit{material}: materials to be used during maintenance that are not chemical elements;
    \item \textit{mode}: ways in which something could be performed;
    \item \textit{measurement unit}: measurement units;
    \item \textit{machine}: typologies of machine (not simple components);
    \item \textit{time}: adverbs or expressions indicating frequencies, periods etc.;
    \item \textit{security}: collective (not personal) security devices;
    \item \textit{user}: categories of users;
    \item \textit{company}: company names;
    \item \textit{standard}: laws, UNI, CEN and ISO standards, legislative decrees etc.
\end{itemize}

New entities distribution among classes is shown in Table \ref{tab:tab5}.

\begin{table}[H]
\centering
\caption{Distribution of new entities among classes}
\label{tab:tab5}
\begin{tabular}{ccc}
\hline
\textbf{label} & \textbf{\begin{tabular}[c]{@{}c@{}}number of \\ new entities\end{tabular}} & \textbf{proportion} \\ \hline
blacklist      & 258                                                                        & 27,65\%             \\
operation      & 160                                                                        & 17,15\%             \\
component      & 152                                                                        & 16,29\%             \\
verb           & 138                                                                        & 14,79\%             \\
material       & 45                                                                         & 4,82\%              \\
machine        & 39                                                                         & 4,18\%              \\
mode           & 39                                                                         & 4,18\%              \\
chemistry      & 37                                                                         & 3,97\%              \\
measure        & 18                                                                         & 1,93\%              \\
drawbacks      & 10                                                                         & 1,07\%              \\
time           & 10                                                                         & 1,07\%              \\
advantages     & 7                                                                          & 0,75\%              \\
utensile       & 7                                                                          & 0,75\%              \\
qualifica      & 4                                                                          & 0,43\%              \\
security       & 4                                                                          & 0,43\%              \\
user           & 2                                                                          & 0,21\%              \\
company        & 1                                                                          & 0,11\%              \\
DPI            & 1                                                                          & 0,11\%              \\
standard       & 1                                                                          & 0,11\%              \\ \hline
\textbf{Total} & \textbf{933}                                                               & \textbf{100\%}      \\ \hline
\end{tabular}
\end{table}

The new entities were also automatically associated to their frequency, representing the number of times the same entity was identified in the relevant sentences. Table \ref{tab:tab6} shows an extract of new entities and associated classes and frequencies.

\begin{table}[H]
\centering
\caption{Extract from new entities and associated classes and frequencies}
\label{tab:tab6}
\begin{tabular}{ccc}
\hline
\textbf{frequency} & \textbf{new entity} & \textbf{class} \\ \hline
30                 & provvedere          & verb           \\
26                 & caso                & blacklist      \\
24                 & macchina            & machine        \\
21                 & verificare          & verb           \\
18                 & stato               & blacklist      \\
17                 & lubrificazione      & operation      \\
11                 & pulizia             & operation      \\
10                 & manutenzione        & operation      \\
9                  & gruppo stampa       & component      \\
9                  & sostituzione        & operation      \\
9                  & contattare          & verb           \\
8                  & trasmissione        & operation      \\
7                  & gruppo              & blacklist      \\
7                  & aspiratore          & machine        \\
7                  & fustellatura        & operation      \\
7                  & necessità           & blacklist      \\
6                  & prima               & blacklist      \\
6                  & operazioni          & blacklist      \\
6                  & sistema             & blacklist      \\
6                  & aggressivo          & blacklist      \\
...                & ...                 & ...            \\ \hline
\end{tabular}
\end{table}

\section{Final Results}

The most relevant result of this research is not only the automatic extraction of relevant entities from the maintenance manual, but also the definition of their relations, because a chatbot engine requires them in order to manage a conversation.

The correlation and interconnection of entities contained within the taxonomy is explained through Network Analysis, quantified by measuring their co-occurrence in BOBST manual sentences. Building a graph based only on the co-occurrence measure is the first step towards the construction of a more complex network of relations (such as lexical or semantic similarities). Furthermore, the graph is the best way of representing and visualizing intuitively the output of the whole process.

Once we collected the entities and quantified the relation among them, we represented this structure as a graph where we can find:

\begin{itemize}
    \item a set of nodes, representing the entities, whose attributes are:
    \begin{itemize}
        \item size: the absolute frequency in the manual;
        \item colour: the class of the entity;
        \item label: the name of the entity
    \end{itemize}
    \item a set of edges, representing the relation between entities: the higher the thickness, the stronger the relation.
\end{itemize}

\begin{figure}[H]
  \centering
  \includegraphics[keepaspectratio=true,scale=0.4]{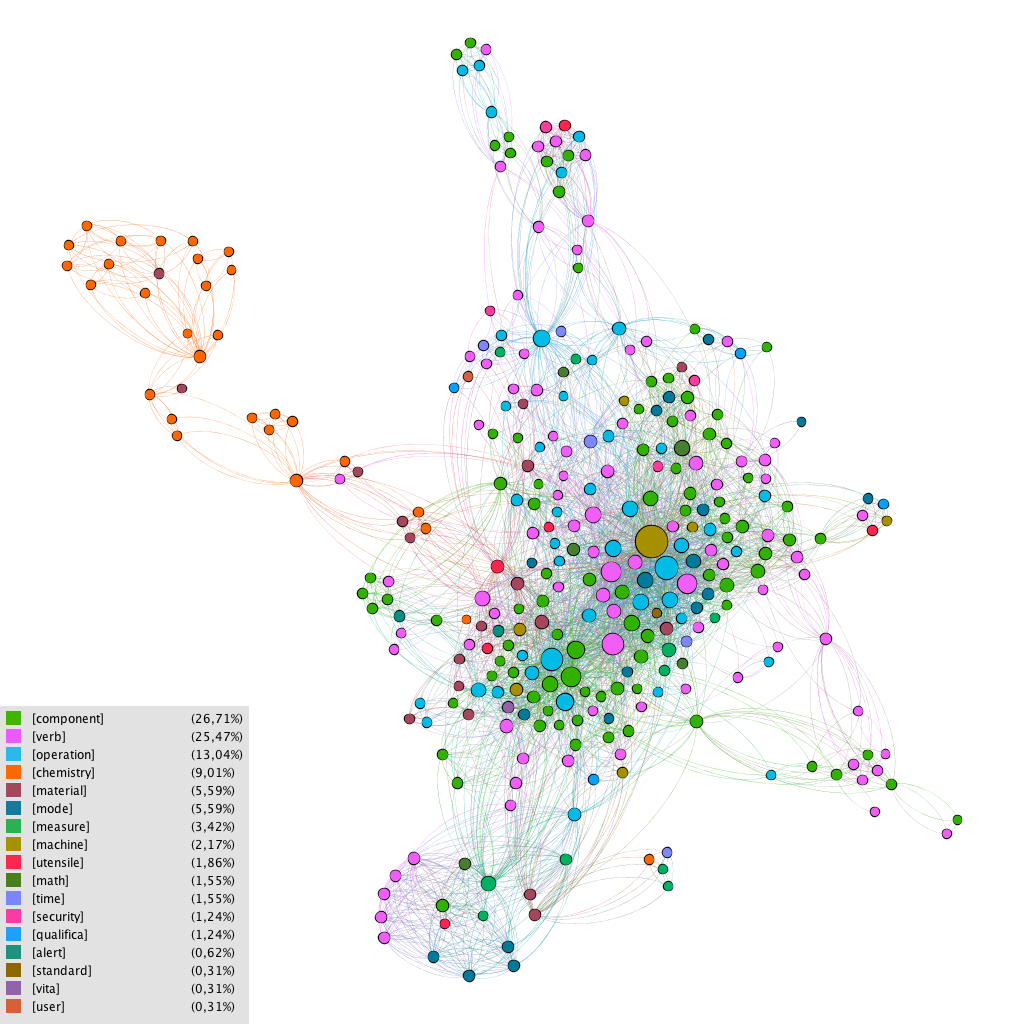}
  \caption{Taxonomy network of BOBST manual}
  \label{fig:fig3}
\end{figure}

The taxonomy network is shown in Figure \ref{fig:fig3}.

\begin{figure}[h!]
  \centering
  \includegraphics[keepaspectratio=true,scale=0.4]{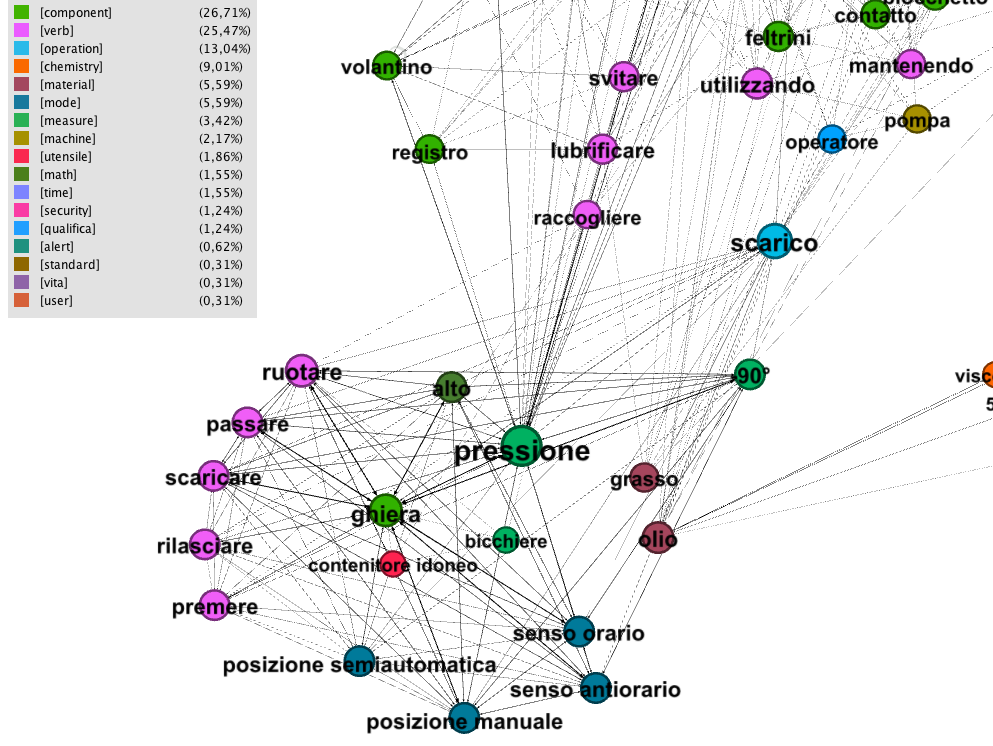}
  \caption{Detailed extract of the taxonomy network}
  \label{fig:fig4}
\end{figure}

As could be seen in Figure \ref{fig:fig3}, many entities are strongly linked, making the results of the research very interesting to be interpreted. Analyzing the network, the relations among the entities could be deeply explored. In order to make the reader understand the potential of the network, we decided to show in detail an extract of the graph in Figure \ref{fig:fig4}.

The lower part of Figure \ref{fig:fig4} shows some entities strongly linked: all these have in common the component ghiera (the ring). In purple there are 5 verbs that are linked to the component in the manual: this means that the ring could be for example rotated (ruotare), or pressed (premere). These operations should be performed using a certain pressure, pressione (the unit of measure in pink) and following different ways (for example the ring could be rotated in manual position, posizione manuale, colored in blue).
The upper part of Figure \ref{fig:fig4} shows 3 verbs (\textit{svitare}, to unscrew, \textit{lubrificare}, to lubricate, and \textit{raccogliere}, to pick up) representing actions performed on two very similar components (the valve, v\textit{volantino}, and the regulator, \textit{registro}).

The strength of the system is that it is able to generate different graphs for different sources and this result leads to a first attempt of knowledge representation of the maintenance process. Although this is specific for the BOBST case, it could capture a broader view of the domain we are dealing.

\section{Conclusions and Future Developments}

This research highlights the first step towards automatic building of a chatbot to support maintenance operations by automatically extracting, from technical documents, entities and their relations and mapping them in a taxonomy network. The lack of human intervention makes the process scalable, enabling, for example, the support for other business functions. Further applications on a large number of maintenance manuals will enlarge the body of knowledge, improving the entity extraction and providing more accurate relations. 

The commitment of a multinational company makes the contribution to practice relevant since the methodology is applied on a real case, based on practical needs and makes chatbots development faster and cost-efficient. For what concerns the contribution to scholarship, it is mainly identifiable in the improvement of every phase of the knowledge management, proved to be a crucial topic in Industry 4.0 environment.

The application of the method led us to face different issues and, consequently, to reflect upon the weaknesses of the approach that need to be improved.
For this application, it has been decided not to stem (i.e. transform into the root form) the entities belonging to the expanded taxonomy. Although, the limitations concerning the italian language entail more complex considerations in future works, such as stemming and PoS tagging. For example, since stemming (the task of extracting the root of a word) is useful to avoid problems linked to plural, gender and verb inflections, the large number of derived forms in the italian language could wrongly merge two words that originally have different meanings (for example, vita (the life) and vite (the screw) whose stem is “vit”).
Furthermore, the classes of entities need to be organized in different hierarchical levels and the use of lexical pattern will improve the extraction task: for example, UNI, CEN and ISO standard are always expressed by following the same lexical pattern; the string UNI (or CEN or ISO) is always followed by a string of three, four or five numeric characters, the two strings could be consecutive or separated by a blank space.

Although the proposed methodology is a first approach to the automatic building of a chatbot mining information from technical documentation, it allowed us to achieve a good result. We are confident that the approach could be improved starting from the previous considerations.
The next step of the research will be the building of a domain-specific knowledge base for BOBST maintenance operations.

\section*{Acknowledgement}
The authors are very thankful to the company BOBST SA for the material provided and for its support to the research.

\bibliographystyle{unsrt}  



\end{document}